\documentclass[aps,prl,preprint,groupedaddress]{revtex4}
\usepackage{ifthen}
\usepackage{tikz}
\usepackage{amsmath,amssymb}
\usepackage{graphicx}
\usepackage{wrapfig}
\usepackage{subfig}
\usepackage[squaren]{SIunits}
\addunit{\molar}{M}
\addunit{\calory}{cal}
\addunit{\eu}{e.u.}
\usepackage{booktabs}
\usepackage{dcolumn}

\newcommand\be{\begin{equation}}
\newcommand\ee{\end{equation}}

\newcommand\Kt[1]{\tilde\kappa\ifthenelse{\equal{#1}{}}{}{({#1})}}

\newcommand\Zggt[2]{
  \mathcal{\tilde{Z}}
  \ifthenelse
  {\equal{#1}{}}
  {}
  {({#1},{#2})}
} 
\newcommand\Zgt[2]{{{\tilde{Z}}_{#1}\ifthenelse{\equal{#2}{}}{}{({#2})}}} 

\begin{document}


\title{TT2NE: A novel algorithm to predict RNA secondary structures with pseudoknots}


\author{Micha\"el Bon}
\author{Henri Orland} \affiliation{Institut de Physique
Th\'eorique, \\ CEA Saclay, CNRS URA 2306, \\ 91191 Gif-sur-Yvette, France}


\date{\today}

\begin{abstract}
We present TT2NE, a new algorithm to predict RNA secondary structures with pseudoknots. The method is based on a classification of RNA structures according to their topological genus. TT2NE guarantees to find the minimum free energy structure irrespectively of pseudoknot topology. This unique proficiency is obtained at the expense of the maximum length of sequence that can be treated but comparison with state-of-the-art algorithms shows that TT2NE is a very powerful tool within its limits. Analysis of TT2NE's wrong predictions sheds light on the need to study how sterical constraints limit the range of pseudoknotted structures that can be formed from a given sequence. An implementation of TT2NE on a public server can be found at http://ipht.cea.fr/rna/tt2ne.php.
 \end{abstract}


\maketitle

\section{Introduction}
In the past twenty years, there has been a tremendous increase of
interest of the biological community for RNA. This biopolymer,
which was at first merely considered as a simple information
carrier, was gradually proven to be a major actor in the biology
of the cell \cite{book}. It was first discovered that some RNAs
might have enzymatic activity (ribozymes) and as such would
directly play a crucial role in the biochemical reactions taking
place in the cell. More recently, it was also discovered that some
RNAs, in particular micro-RNAs, have a post-transcriptional
regulation role in the cell by controlling the level of
translation of some messenger RNAs. Up to 30\% of human genes
might be regulated by such micro-RNAs. At present, it is also
believed that a considerable amount of ``junk" (non-coding) DNA is
transcribed into some non-coding RNAs, the role of which is still
unclear.

Since the RNA functionality is mostly determined by its three-dimensional
conformation, the accurate prediction of RNA folding from
the base sequence is a central
issue \cite{Tinocco1991}.
It is strongly believed that the biological activity of RNA (be it enzymatic or regulatory), is implemented through the binding of some unpaired bases of the RNA with their ligand. It is thus crucial to have a precise and reliable map of all the pairings taking place in RNA and to correctly identify loops.
The complete list of all Watson-Crick and Wobble base pairs in RNA is called the {\em secondary structure} of RNA.

Since the folding of even short RNA molecules takes too
long to perform with all-atoms simulations including explicit solvent,
the more modest goal of solely obtaining the most probable secondary structures
based on experimentally derived base-pairing and base-stacking free energies
has been pursued.
It seems very plausible that (as in NMR protein structure prediction) the secondary structure of RNAs is sufficiently constraining to entirely and unambiguously determine the 3-dimensional structure of the molecule. This 3-dimensional structure of the RNA in turn controls the biochemistry of the molecule, by making certain regions of its surface accessible to the ligand molecule.

In this paper, we will adhere to the notion that there is an
effective free energy which governs the formation of secondary
structures, so that the optimal folding of an RNA sequence is
found as the minimum free energy structure (MFE for short). The
problem of finding the MFE structure given a certain sequence has
been conceptually solved provided the MFE is planar, \emph{ie} the
MFE structure contains no pair ($i$,$j$), ($k$,$l$) such that
$i<k<j<l$. In that case, polynomial algorithms which can treat
long RNAs assuming a mostly linear free energy model have been
found \cite{nussinov,zuker0,mccaskill}. Otherwise, the MFE
structure is said to contain pseudoknots and finding it has been
shown to be an NP-complete problem with respect to the sequence
length \cite{lyngso2}. Even if pseudoknots represent a small part
of known structures, they often have a functional role \cite{kuo,
pleijframe} and the problem of their prediction must be addressed.

Three main algorithmic strategies can be thought of to take into account the NP-completeness of pseudoknotted MFE prediction : 1) empirical search of the MFE using heuristic methods,  2) efficient exact calculations on a restricted class of pseudoknots  and 3) exact calculations, using various tricks to allow for the treatment of as long as possible sequences.


Here we present TT2NE, an algorithm that falls into the latter category. TT2NE relies on the ``maximum weighted independent set" (WIS) formalism. In this formalism, an RNA structure is viewed as an aggregate of stem-like structures (helices or helices comprised of bulges of size 1 or internal loops of size $1 \times 1$). These stem-like structures can be viewed as points in the space of all helical fragments available from a given sequence and we will refer to them as ``helipoints". Please note that our notion of helipoints is in fact not trivial and differs from what is done in algorithms based on the WIS formalism, where they generally reduce to maximum helices (see the explanation in material and methods).
Given a certain sequence, the set of all possible helipoints is computed and a weighted graph is built in the following way:
\begin{itemize}
\item the vertices of the graph are the helipoints, with a weight given by the opposite of their free energy of formation,
\item two vertices are connected by an arch if and only if the corresponding helipoints are not compatible in the same secondary structure.
\end{itemize}
Indeed, two helipoints may be mutually exclusive in a graph: this is for example the case if they share at least one base (since triplexes are forbidden).
Finding the MFE structure thus amounts to finding the maximum weighted independent set of the graph, i.e. the set of pairwise compatible helipoints such that the overall free energy is minimum.

\begin{figure}
\centering
\includegraphics[scale=0.5]{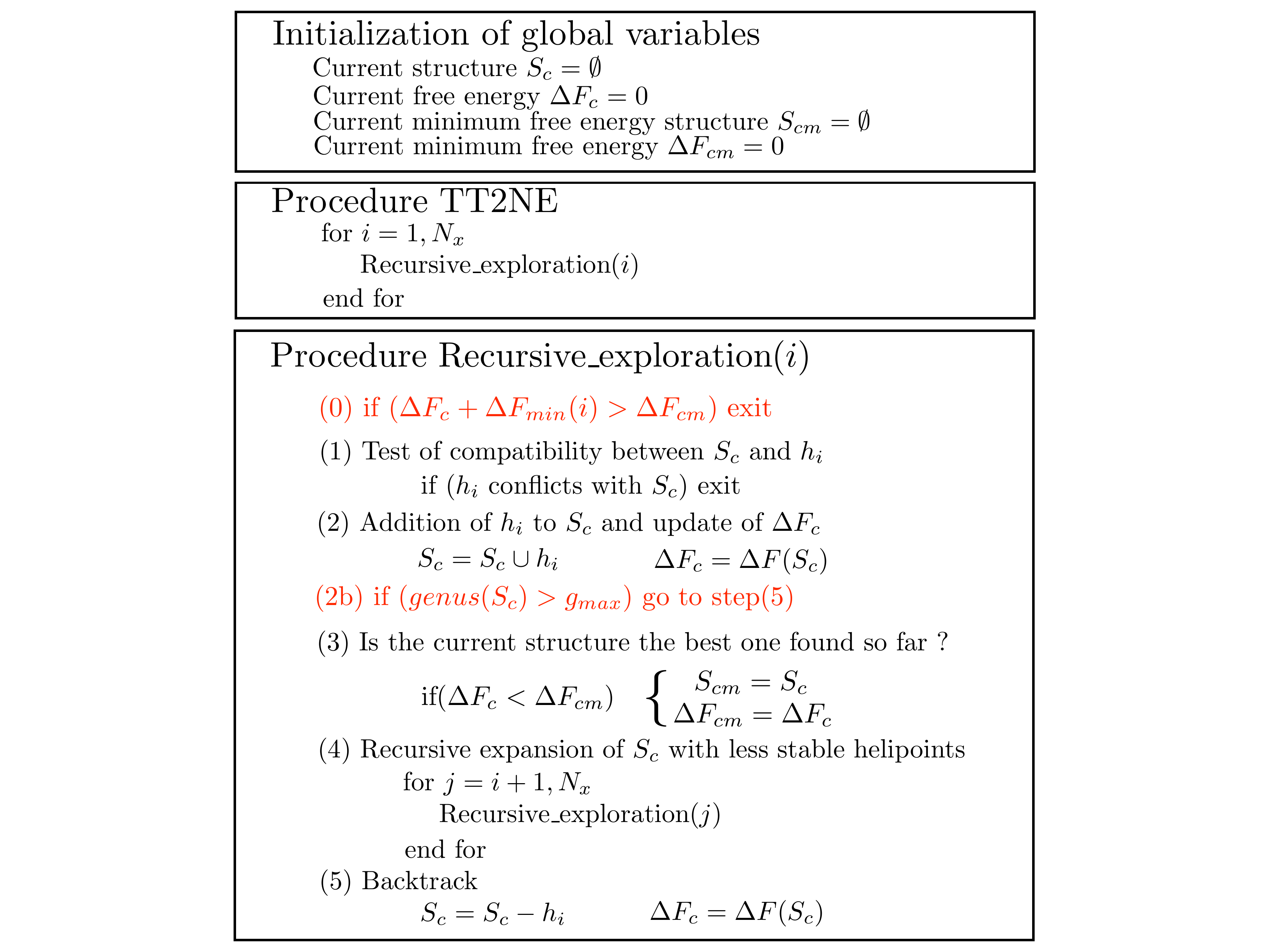}
\caption{Pseudocode of TT2NE. \emph{The base routine is written in black and performs an exhaustive enumeration of all independent sets of $\mathcal G_{x}$. In the end, the MFE structure can be read in the global variable $\Delta F_{cm}$. The two red lines are improvements discussed in the text.}}
\label{pc}
\end{figure}

Given a certain sequence $x$, let's note $N_{x}$ the number of available helipoints and $\mathcal G_{x}$ the associated graph.
The base routine of TT2NE is a simple exhaustive depth exploration of all independent sets of $\mathcal G_{x}$ using a backtracking procedure, where vertices are added to the current structure in the increasing order of their free energy, that is decreasing order of weight (see black pseudocode in Fig. \ref{pc}). There is in particular no restriction on the pseudoknots topologies that TT2NE can generate. However, this strategy is very inefficient. In this article we propose two ideas to improve it. First, we use a new treatment of pseudoknots that restrain TT2NE's search to a much smaller and relevant subspace of independent sets. Second, we take advantage of a peculiar energy model to enforce a branch-and-bound procedure that speeds up the search of the MFE without loss of exactness. A server implementation of TT2NE can be found at http://ipht.cea.fr/rna/tt2ne.php.

\subsection{A new treatment of pseudoknots}
In a previous series of studies \cite{orland,bon}, we have proposed a classification of pseudoknots according to their topological genus. The genus is an integer number that captures the complexity of a pseudoknot and we have shown that naturally occurring pseudoknots have a much lower genus than expected in randomly paired polymers \cite{bon}.
In particular, we have shown that for sequences of sizes up to 500 bases, the genus does not exceed 2. For sizes around 1500 bases, the genus ranges between 2 and 6. Finally, for the largest RNAs (around 3000 bases) the genus may reach 17.

We use this fact to guide TT2NE's search of relevant pseudoknots in two ways. First, a penalty for pseudoknot formation depending on their genus is introduced in the free energy model. Although more sophisticated forms could be imagined, for now we chose a simple linear form.
A pseudoknot of genus $g$ is assigned a penalty $+\mu g$ where we set $\mu$ to $+1.5$ kcal/mol. This value of $\mu$ was obtained by optimizing the number of correctly predicted structures by our algorithm.
Second, an upper limit $g_{max}$ is introduced. This limit, tunable by the user, has a critical importance as it defines the space of pseudoknots where TT2NE will restrain its search. The size of this space grows exponentially with $g_{max}$, so this number has a great impact on the computational time required by TT2NE.  Based on the relation of RNA size to genus mentioned above, we may safely fix a maximum genus of 3 for RNA sizes smaller than 250, typically the maximal size we can treat with our present algorithm due to computational time constraints.

We have shown that the most standards pseudoknots, i.e the H-pseudoknot and the kissing-hairpin, have both genus 1. It implies that if one is interested in short chains which carry these kind of pseudoknots, setting $g_{max}$ to 1 is sufficient and would save a lot of computational time. Setting $g_{max}$ to a large value would leave the problem as open as possible, but again, a wise tuning of this parameter proves a relevant and efficient way to locate the MFE in a fast way.

\subsection{A branch-and-bound procedure}
The base routine of TT2NE can be improved using a branch-and-bound procedure. The idea is to speed up the search of the MFE of $\mathcal G_{x}$ by computing first the MFE of some relevant subgraphs. The crux of such a branch-and-bound procedure is to be able to relate those partial solutions to the general problem and this can be done in TT2NE by taking advantage of a peculiar energy model.

\subsubsection{Energy model}
Vertices are sorted in increasing order of free energy, ie the vertex 1 represents the most favorable helipoint. We note  $\Delta F_{i}$ the free energy of the $i^{th}$ vertex. Then in TT2NE the free energy of a structure S made of helipoints $\{ h_{i} \}_{i \in \Omega(S)}$ is computed with the following model $M_{1}$ :
\begin{equation}
\Delta F^{M_{1}}(S) = \sum_{\substack{i \in \Omega(S)}} \Delta F_{i} + \nu_{m} n_{m}(S) + \mu g(S) \label{M1}
\end{equation}
where $n_{m}(S)$ is the number of multibranch loops of $S$ and $\nu_{m}$ is the corresponding penalty of formation. Note that in this model there is no term for large internal loops or bulges. We also introduce the simple model $M_{0}$ where the free energy of $S$ is just the sum of the free energies of the helipoints it is made of :
\begin{equation}
\Delta F^{M_{0}}(S) = \sum_{\substack{i \in \Omega(S)}} \Delta F_{i}
\end{equation}
\subsubsection{Property}
Let $\Delta F_{min}(i)$ be the MFE of structures comprised of helipoints with indices larger than $i$, according to the energy model $M_{0}$. $\Delta F_{min}(i)$ would simply be the output of TT2NE when used on the restriction of $\mathcal G_{x}$ to its $N_{x}-i$ last vertices with model $M_{0}$. Let $S^{0}$ be a structure made of $n$ helipoints  and $i_{n}$ the index of its least stable helipoint. Let's note $S_{/k}$ the restriction of a structure S to its $k$ most stable helipoints. Then it can be straightforwardly shown that the following property holds :
\begin{equation}
\forall S, \ \ S_{/n}=S^{0}  \Rightarrow \  \Delta F^{M_{j}}(S) \ge  \Delta F^{M_{j}}(S^{0}) +  \Delta F_{min}(i_{n}+1)  \ \ \mbox{for} \ j=0 \ \mbox{or} \ 1 \label{bab} \\
\end{equation}

The practical meaning of this relation is : there is a lower limit to the free energy of all structures that can be derived from $S^{0}$ by adding any combination of helipoints of indices more than $i_{n}$. Consequently, if this lower limit is found to be larger than the current MFE that TT2NE has found so far, TT2NE can safely ignore all these structures : the global MFE cannot be found in this ensemble. This property thus allows to further restrain the size of the search space for the MFE.

Those two improvements can be incorporated in TT2NE as can be seen in red in Fig. \ref{pc}.

\section{Materials and methods}

\subsubsection{Efficient calculation of the genus}
\begin{figure}[htp]
\centering
\subfloat[$P=5,L=5 \rightarrow g=0$] {\includegraphics[scale=0.22]{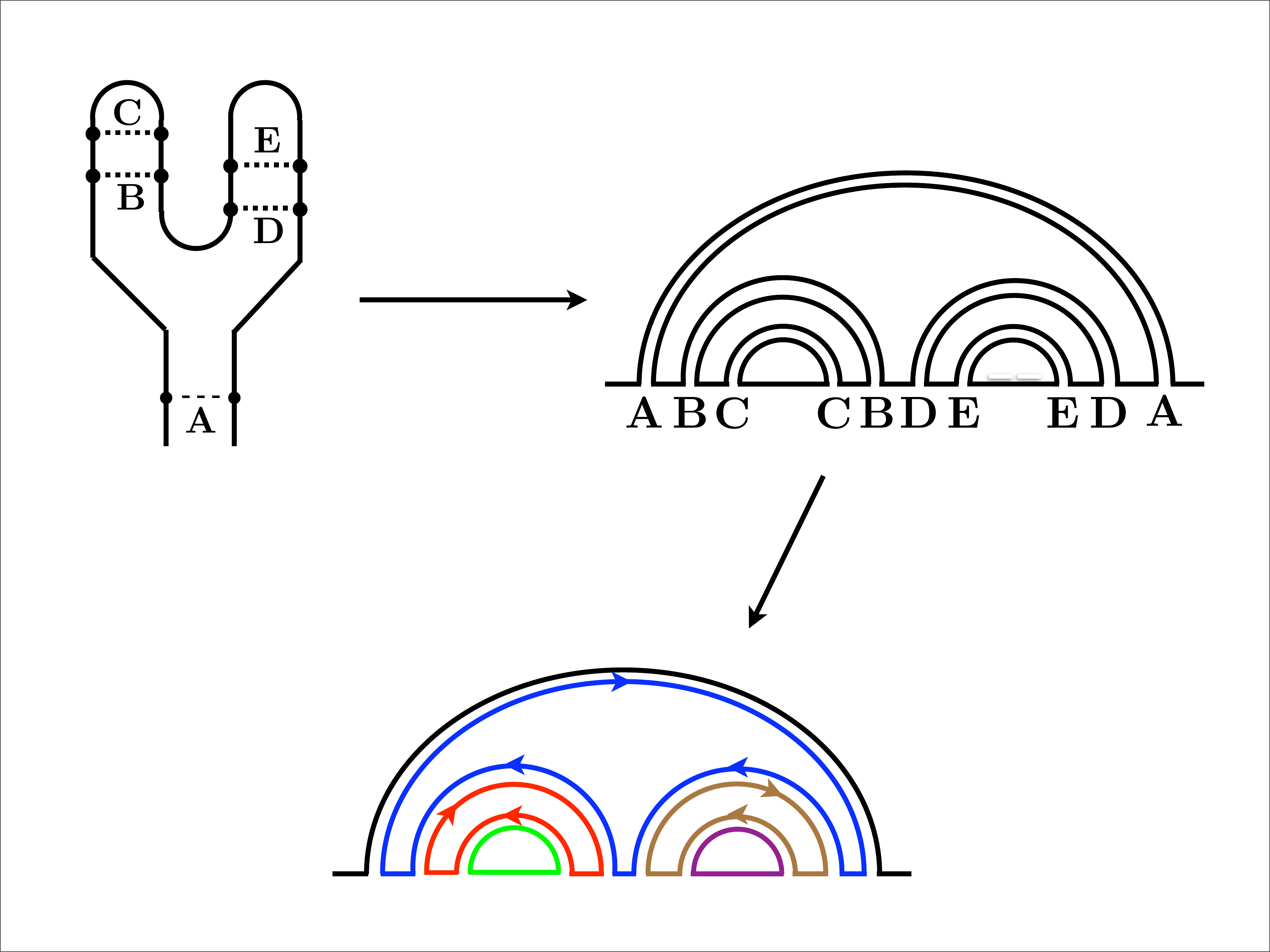}}
\subfloat[$P=5,L=3 \rightarrow g=1$] {\includegraphics[scale=0.22]{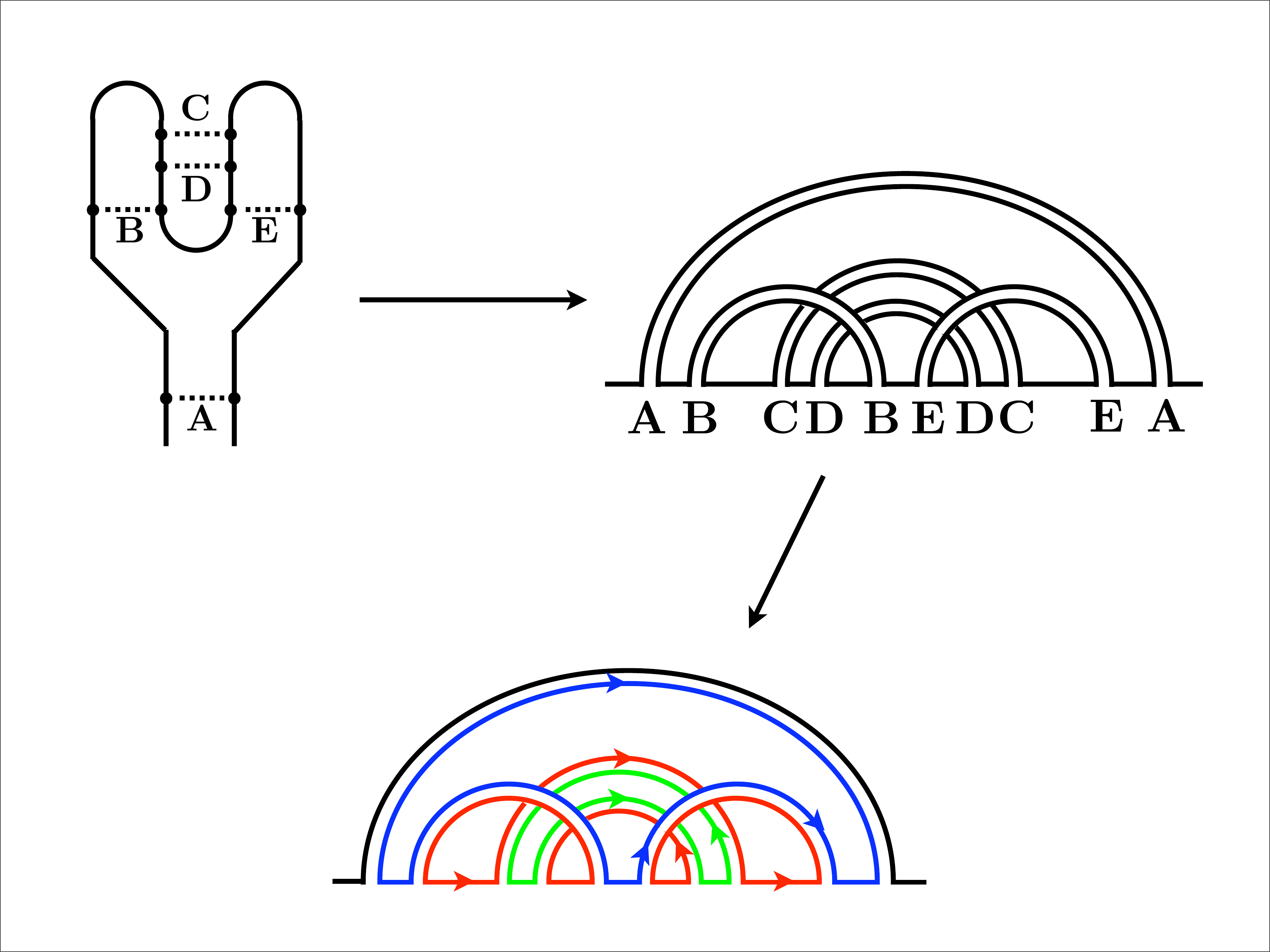}}
\caption{\emph{Examples of how to calculate the genus with a double-line diagram representation.}}
\label{ex_genre}
\end{figure}

TT2NE requires to be able to efficiently update the genus of a structure upon addition or removal of a helipoint.  In order to do so, we use a technique which was introduced by t'Hooft \cite{t}. A structure of RNA is represented as a diagram whose arches are double lines that connect paired bases, such as represented in Fig.m \ref{ex_genre} .

In this process, loops are created within those diagrams and it can be shown that the genus of the corresponding structures can simply be calculated with :
\be
g=\frac{P-L}{2}
\ee
where $P$ is the number of pairs and $L$ the number of loops. Upon addition of a new pair to a structure, the genus variation $\Delta g$ is given by
\be
\Delta g=\frac{1-\Delta L}{2}
\ee
We found a property that allows to calculate the term $\Delta L$ in an efficient way. Upon addition of a pair (i,j) to a certain diagram,

\be
\Delta L = \left\{
\begin{array}{rl}
1 & \mbox{if $i$ and $j$ belongs to the same loop} \\
-1 & \mbox{otherwise}\\
\end{array}
\right.
\ee

Therefore, $\Delta g$ can be straightforwardly calculated by checking whether the newly paired bases belong to the same loop and this operation can be efficiently performed in a time linear in the number of pairs of the diagram. The case of the removal of a pair is symmetric.

\subsubsection{Generation of the initial graph}

A helipoint is \emph{an ensemble} of helices that share the same extremal pairs. Given two extremal pairs $(i,j)$ and $(k,l)$, the set $\omega^{ij}_{kl}$ of all helices that end with these two pairs can be generated and their individual energies calculated according to a given energy model. The free energy $F^{ij}_{kl}$ of the helipoint is then computed as

\be
\exp{(-\beta  F^{ij}_{kl})} = \sum_{\substack{h \in \omega^{ij}_{kl}}} \exp{(-\beta E(h))} \ \ \ \ \ \ \ \mbox{with} \ \beta = (k_B T)^{-1} \\ \label{zob}
\ee

Helipoints are stem-like structural building blocks which account for all possible internal pairing possibilities that occur between their extremal pairs. The importance of this notion is well captured by considering for example such a  sequence : GGGAGGG [...] CCCUUCCC. As one can see, a helix containing a ``bulged" uracil can be formed from this sequence, but there are two ways to choose the ``bulged" uracil. In order to describe this fact appropriately in statistical mechanics, it is important nor to neglect any of these possibilities neither to consider them as distinct competitors. Rather, the notion of helipoint implies that both possibilities would \emph{stabilize}  the pairing of these regions of the sequence. In this example, the calculation of the free energy according to equation \ref{zob} would indeed introduce an entropic bonus of $-k_BT \ln 2$ that accounts for this variability. \\

The computation of helipoints free energies requires the setting of some values for the basic structural elements of RNA folds : stacking, terminal mismatches, helix formation penalty, bulges and internal loops. The three first families of terms have been taken from \cite{mathews1999}. We computed the free energy of the bulges of size one as the energy of the stack of pairs closing this bulge plus 3.8 kcal/mol. The energy of a helix comprising a $1 \times 1$ internal loop is computed as the sum of the free energies of the two helices delimited by this internal loop minus 3.85 kcal/mol. Larger internal loops and bulges of size more than one were not taken into account. In particular, helipoints do not include such kind of motifs. The multibranch loop formation penalty was not used (ie set to 0) in the work presented here, even though TT2NE could handle it. All helipoints of favorable (ie negative) free energies were kept to build the graph. Note that in most other algorithms based on the WIS formalism, only \emph{maximal} favorable helices are kept (i.e. helices such that the outer nearest neighbors of their extremal pairs cannot pair). Our choice not to restrict our algorithm to maximal helipoints makes the problem harder since it makes the graph wider, but the reason will be explained in the discussion part below.\\
Two helipoints were considered incompatible (i.e. they are connected in the graph) if :
\itemize{
\item  they overlap
\item their concatenation generates an existing helipoint.
\item their concatenation produces a sterically impossible structure.
} \\

This last requirement anticipates on a point that will be explained in the ``discussion" section.

\subsubsection{Branch-and-bound procedure}
The equation \ref{bab} requires a prior computation of the terms $\Delta F_{min}(i)$, that is the MFE of $\mathcal G_{x}$ restrained to helipoints of index larger than $i$. Those quantities are obtained by running TT2NE on those subgraphs. However, calculating those terms for all $i$ is useless  since the only needed quantity is  $\Delta F_{min}(1)$. Rather, one must choose a certain level up to which these terms should be calculated, in order to get a good balance between the time spent in doing so and the time saved later in the search of the MFE. In the work presented here, we generally computed the quantities $\Delta F_{min}(i)$ for the $350$ least stable helipoints.

\subsubsection{Suboptimal structures}
The algorithm presented here only outputs the MFE. It is very easy to adapt it to instead output  a certain number of suboptimal structures specified by the user if needed.

\subsubsection{Heuristic}
For longer sequences, a heuristic can be used : the above techniques are first applied to the restriction of the graph to its $N_{h}$ most stable helipoints and the best structures output are then saturated with the remaining helipoints. This heuristic is identical to the initial problem with $N_{h}=N_{x}$ and becomes more and more imprecise as $N_{h}/N_{x} \rightarrow 0$.

\subsection{Detailed results}

We compared TT2NE with McQfold \cite{Metzler}, HotKnots \cite{ren} and Mfold \cite{mfold} on a set of 35 sequences which is quite similar to the set used in the original HotKnots paper. We did not compare it with the Pknots algorithm of Rivas and Eddy \cite{eddy} as its computation time is very long (it scales like the 6th power of the length of the sequence). Sequences were mostly retrieved from the Pseudobase \cite{pseudobase} and are named after their Pseudobase entry with the exception of the sequence ``1u8d" which is named after its PDB entry. For each sequence, sensitivity and positive predicted value (PPV) have been measured. The sensitivity is defined as the fraction of correctly predicted pairs of the native structure.  The PPV is defined as the fraction of correctly predicted pairs of the predicted structure. Both are indicated in \% in the following array (see Table 
below). Stars are pointing to sequences where the correct structure is actually the second best prediction. For each sequence, the best sensitivity predicted is emphasized in boldface. In all those tests, TT2NE's parameter $g_{max}$ was set to $3$.\\

\begin{tabular}{|c|c|c|c|c|c|c|c|}
\hline
sequence&length&genus&Mfold&HotKnots&McQfold&TT2NE&genus TT2NE  \\
\hline
1u8d & 68 &1& 69 - 100 & 69 - 100 & 69 - 100 &\textbf{88} - 100&1 \\
\hline
 AMV3&113 &1&84 - 86 &84 - 86 & 76 - 81 &\textbf{87} - 85&1  \\
\hline
 BBMV &116&1& 0 - 0 & 81 - 81&  \textbf{86} - 82 & \textbf{86} - 84&1  \\
 \hline
 Bp\_PK2 &91&1&81 - 96& 81 - 96& 87 - 87 &\textbf{100} - 100 &1\\
 \hline
 BVDV &74&1&52 - 65& 52 - 61 & 76 - 82 & \textbf{96} - 96&1 \\
\hline
 BWYV &51&1&55 - 55&  \textbf{100} - 69  & 55* - 55 & \textbf{100} - 100&1 \\
\hline
 Bt-PrP&45&1&41 - 33& 41 - 38 & \textbf{50} - 40 &\textbf{50} - 35&1 \\
\hline
CcTMV &73&3& 23 - 27&23 - 27 & \textbf{57} - 93 &42 - 52&0\\
\hline
CGMMV &85&3& 58 - 69&\textbf{67} - 87&38 - 48 & 58 - 72&0  \\
\hline
 CoxB3 &73&1& 68 - 89&68 - 89 & \textbf{92} - 100 & \textbf{92} - 100&1 \\
\hline
 Ec\_alpha &108&1& 45 - 29&45 - 29 & 50 - 37 & \textbf{79} - 61 &1 \\
\hline
 Ec\_PK1 &31&1&0 - 0 & \textbf{100} - 90 & \textbf{100} - 90 & \textbf{100} - 90&1\\
  \hline
 EC\_PK4 &52&1& 0 - 0&68 - 100 &52 - 71 & \textbf{100} - 100 &1\\
\hline
 Ec-RpmI &72&1& \textbf{68} - 90&20 - 26 & 51 - 71 & 58 - 60&1 \\
\hline
 Ec\_S15&67&1& 58 - 62&\textbf{100} - 73 & 58* - 62 & \textbf{100} - 73&1 \\
\hline
 GLRaV-3 &75&1&65 - 59& 65 - 59 & \textbf{100} - 76 &\textbf{100} - 76&1 \\
\hline
 HAV &55&1&  \textbf{58} - 83& \textbf{58} - 83 &  \textbf{58} - 83 & \textbf{58}* - 83&0\\
\hline
 HCV\_229E &74&1& 79 - 100&79 - 100 &\textbf{100} - 100&100 - 100&1 \\
\hline
 HDV &87&2& 65 - 70&41* - 44 &75 - 75&\textbf{93} - 84&2\\
\hline
 HDV\_anti &91&2& 16 - 14&16* - 14 &\textbf{100} - 80&72 - 58&2 \\
\hline
 Hs\_PrP&45&1& 0 - 0 & 0- 0 &\textbf{54} - 42& 0-0 & 0 \\
\hline
 IBV &56&1& 55 - 66&\textbf{100} - 100 &94 - 100&94 - 100&1 \\
\hline
 Lp\_PK1&31&1& \textbf{50} - 100&\textbf{50}* - 100 &\textbf{50} - 100&\textbf{50}* - 100&0\\
\hline
\end{tabular}

\begin{tabular}{|c|c|c|c|c|c|c|c|}
\hline
sequence&length&genus&Mfold&HotKnots&McQfold&TT2NE&genus TT2NE  \\
\hline
 Mengo-PKC&26&1& 37 - 60& 0 -  0 & 37 - 60&\textbf{100} - 100&1 \\
 \hline
 minimalIBV&45&1& 64 - 91&\textbf{100} - 94& \textbf{100} - 94&\textbf{100} - 94&1 \\
\hline
 MMTV&34&1& 0 - 0 & \textbf{100} - 91 &\textbf{100} - 91 &\textbf{100} - 91&1 \\
\hline
 pKA-A&36&1& 50 - 66&\textbf{100} - 92&\textbf{100} - 92&\textbf{100} - 92&1 \\
 \hline
 RSV&128&1& 74 - 76&97 - 82 &\textbf{100} - 95&94 - 88&1\\
 \hline
 satRPV&73&1& 59 - 68&59 - 68&\textbf{81} - 81&\textbf{81} - 81&1\\
 \hline
 SRV-1&38&1& 0-0 & \textbf{100} - 100&\textbf{100} - 100&\textbf{100} - 100&1 \\
\hline
 T2\_gene32&33&1& 58 - 70&\textbf{100} - 100& \textbf{100} - 100 & \textbf{100} - 100&1\\
\hline
 T4\_gene32&28&1& 63 - 87&63* - 87&63 - 100&\textbf{100} - 100&1 \\
\hline
 TMV &74&3& \textbf{52} - 65&\textbf{52} - 61&\textbf{52} - 65&48 - 54&1 \\
 \hline
 Tt-LSU&65&1& 60 - 75&\textbf{95} - 100&60- 100&\textbf{95} - 100&1 \\
\hline
 TYMV&74&1& \textbf{72} - 78&70 - 73& \textbf{72} - 78& \textbf{72} - 69&1 \\
\hline
 average& & &54 - 59&65 - 70&75.5 - 80&\textbf{82} - 81& \\
\hline
\end{tabular}

On the average, TT2NE achieves better performances on this set of test sequences.  Comparison with HotKnots shows that these improvements originate from the different treatment of pseudoknots, as HotKnots and TT2NE otherwise use essentially the same energy model. \\

\subsection{Comments and discussion}
Despite the fact that TT2NE can find any type of topology and guarantees to output the MFE, it does not provide a 100\% success. Why is that so? We have investigated the errors generated by TT2NE and we see two main causes: the first relates to the limit of the energy model used and the second is more specific to the nature of pseudoknots.

\subsubsection{Limits due to the free energy model}
The Turner free energy model has been shown to be partly unable to
explain planar secondary structures \cite{mfold_evaluation}. TT2NE
uses only a subset of this model : thus, there are errors coming
from the part of this model we use, and others coming from the
part we do not use.


An example of the first case is provided by the sequence satRPV : the native secondary structure is almost correctly predicted, but  an error is made because the helix
\setlength{\tabcolsep}{0.2mm}
\begin{tabular}{ccccccc}
${}^{2}$ & C&A&G&A&\\
& G&U&C&U&${}^{19}$
\end{tabular} is considered more thermodynamically favorable than the native one
\setlength{\tabcolsep}{0.2mm}
\begin{tabular}{ccccccc}
${}^{1}$ & A&C&A&G&\\
& C&U&G&U&${}^{16}$
\end{tabular}. \\

An example of the latter case can be seen with Ec-RpmI. There, the native structure contains a helipoint containing a $2\times 1$ internal loop.  The thermodynamics properties of $2\times 1 $ internal loops are not properly taken into account in TT2NE. As a consequence, the energy of formation of that helipoint is not found to be negative and therefore it is not recognized as a relevant helipoint to store into the initial graph. In other words, this helipoint is not favorable and is thus not kept in the construction of the graph. This problem could be solved by allowing for the inclusion of $2 \times 1$ internal loops but this would dramatically increase the number of possible helipoints and the running time of TT2NE would grow exponentially.
\subsubsection{Limits due to the absence of steric constraints}

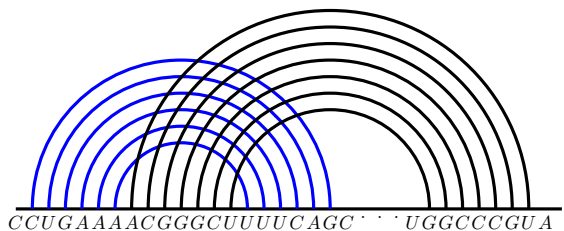
\begin{figure}
\begin{center}
\begin{tikzpicture}

\draw[very thick](0,0) -- (7.26,0);
\draw(0,0) node[below,scale = 0.66]{$C$};
\draw(0.22,0) node[below,scale = 0.66]{$C$};
\draw(0.44,0) node[below,scale = 0.66]{$U$};
\draw(0.66,0) node[below,scale = 0.66]{$G$};
\draw(0.88,0) node[below,scale = 0.66]{$A$};
\draw(1.1,0) node[below,scale = 0.66]{$A$};
\draw(1.32,0) node[below,scale = 0.66]{$A$};
\draw(1.54,0) node[below,scale = 0.66]{$A$};
\draw(1.76,0) node[below,scale = 0.66]{$C$};
\draw(1.98,0) node[below,scale = 0.66]{$G$};
\draw(2.2,0) node[below,scale = 0.66]{$G$};
\draw(2.42,0) node[below,scale = 0.66]{$G$};
\draw(2.64,0) node[below,scale = 0.66]{$C$};
\draw(2.86,0) node[below,scale = 0.66]{$U$};
\draw(3.08,0) node[below,scale = 0.66]{$U$};
\draw(3.3,0) node[below,scale = 0.66]{$U$};
\draw(3.52,0) node[below,scale = 0.66]{$U$};
\draw(3.74,0) node[below,scale = 0.66]{$C$};
\draw(3.96,0) node[below,scale = 0.66]{$A$};
\draw(4.18,0) node[below,scale = 0.66]{$G$};
\draw(4.4,0) node[below,scale = 0.66]{$C$};
\draw(4.62,0) node[below,scale = 0.66]{$.$};
\draw(4.84,0) node[below,scale = 0.66]{$.$};
\draw(5.06,0) node[below,scale = 0.66]{$.$};
\draw(5.28,0) node[below,scale = 0.66]{$U$};
\draw(5.5,0) node[below,scale = 0.66]{$G$};
\draw(5.72,0) node[below,scale = 0.66]{$G$};
\draw(5.94,0) node[below,scale = 0.66]{$C$};
\draw(6.16,0) node[below,scale = 0.66]{$C$};
\draw(6.38,0) node[below,scale = 0.66]{$C$};
\draw(6.6,0) node[below,scale = 0.66]{$G$};
\draw(6.82,0) node[below,scale = 0.66]{$U$};
\draw(7.04,0) node[below,scale = 0.66]{$A$};
\draw[blue,very thick] (0.22,0) arc(180:0:1.98);
\draw[blue,very thick] (0.44,0) arc(180:0:1.76);
\draw[blue,very thick] (0.66,0) arc(180:0:1.54);
\draw[blue,very thick] (0.88,0) arc(180:0:1.32);
\draw[blue,very thick] (1.1,0) arc(180:0:1.1);
\draw[blue,very thick] (1.32,0) arc(180:0:0.88);
\draw[black,very thick] (1.54,0) arc(180:0:2.64);
\draw[black,very thick] (1.76,0) arc(180:0:2.42);
\draw[black,very thick] (1.98,0) arc(180:0:2.2);
\draw[black,very thick] (2.2,0) arc(180:0:1.98);
\draw[black,very thick] (2.42,0) arc(180:0:1.76);
\draw[black,very thick] (2.64,0) arc(180:0:1.54);
\draw[black,very thick] (2.86,0) arc(180:0:1.32);

\end{tikzpicture}

\caption{\emph{Example of a sterically impossible H-pseudoknot}}
\label{steric}
\end{center}
\end{figure}

We also realized that predicting a pseudoknot is not only a question of free energy minimization : steric constraints also matter and some predicted sets of helipoints must sometimes be rejected because they do not correspond to any feasible geometry in 3D space. For example, we display in Fig. \ref{steric} a feature observed in the best secondary structure predicted for the sequence Ec\_alpha (using a standard diagrammatic representation) : \\

This pseudoknot is made of two helices respectively drawn in blue and black. Let's focus on the seven bases of the 5' strand of the black helix (ACGGGCU). The geometry of the nucleotides implies that the pairings organize according to the canonical A-helix shape. However, those seven bases also connect the two ends of the blue helix : they should therefore make up a hairpin loop. It is clear that these two kinds of geometry are mutually exclusive. This diagram therefore cannot match a real RNA structure and must be rejected. To create a sterically allowable pseudoknot between those regions, one or both helices should be shortened. We thus think that a perfect pseudoknot prediction algorithm should be able to include non-maximal helices. This necessity is also very well illustrated by the example of the mouse mammary tumor virus pseudoknot whose 3D structure has been resolved (PDB entry : 1rnk) \cite{1rnk}. This pseudoknot is an H-pseudoknot and one of its helix is non-maximal. By looking at the sequence, one could think that one additional Wobble-pair could form but from looking at the 3D structure, it is clear that due to the peculiar geometry of this pseudoknot, the bases of the putative pair are in fact too far from each other to be able to pair. All algorithms tested on that sequence wrongly predict this additional pair (sensitivity of 1 but PPV of 0.91). We thus have chosen by design to include all possible favorable non-maximal helipoints in the initial graph that TT2NE generates, even though it makes calculations longer.

In fact, it is worth noticing that whenever a pseudoknot is predicted by TT2NE, its PPV is almost always smaller (or equal) than the sensitivity. This means that the predicted structures are somewhat overloaded with spurious pairings. We examined TT2NE's predicted MFE and  we are convinced that most of the time, the helipoints predicted in excess cannot exist due to steric considerations. This point therefore raises an important difference in the evaluation of algorithms for the prediction of secondary structures with and without pseudoknots, such as Mfold. For the latter, if some modifications entails an overall improvement of the sensitivity and the PPV of the predicted MFE, then we can conclude that the predictive power of such an algorithm has been improved. By contrast, with pseudoknot prediction algorithms, such an improvement can be misleading. In fact, the real output  to be taken into account is not the MFE but the first \emph{sterically possible} structure. Even if the predicted MFE has good sensitivity and PPV, it may happen that the best sterically possible structure is in fact completely different and has a bad score. We therefore think that the problem of the determination of sterically impossible structures is essential. As long as we do not know how to detect impossible structures in a fast and efficient way, pseudoknot prediction algorithms may output lots of wrong structures and the evaluation of such algorithms with standard statistical estimates such as sensitivity and PPV of the MFE is quite meaningless.

\begin{figure}[htp]
\begin{center}
\includegraphics[scale=0.3]{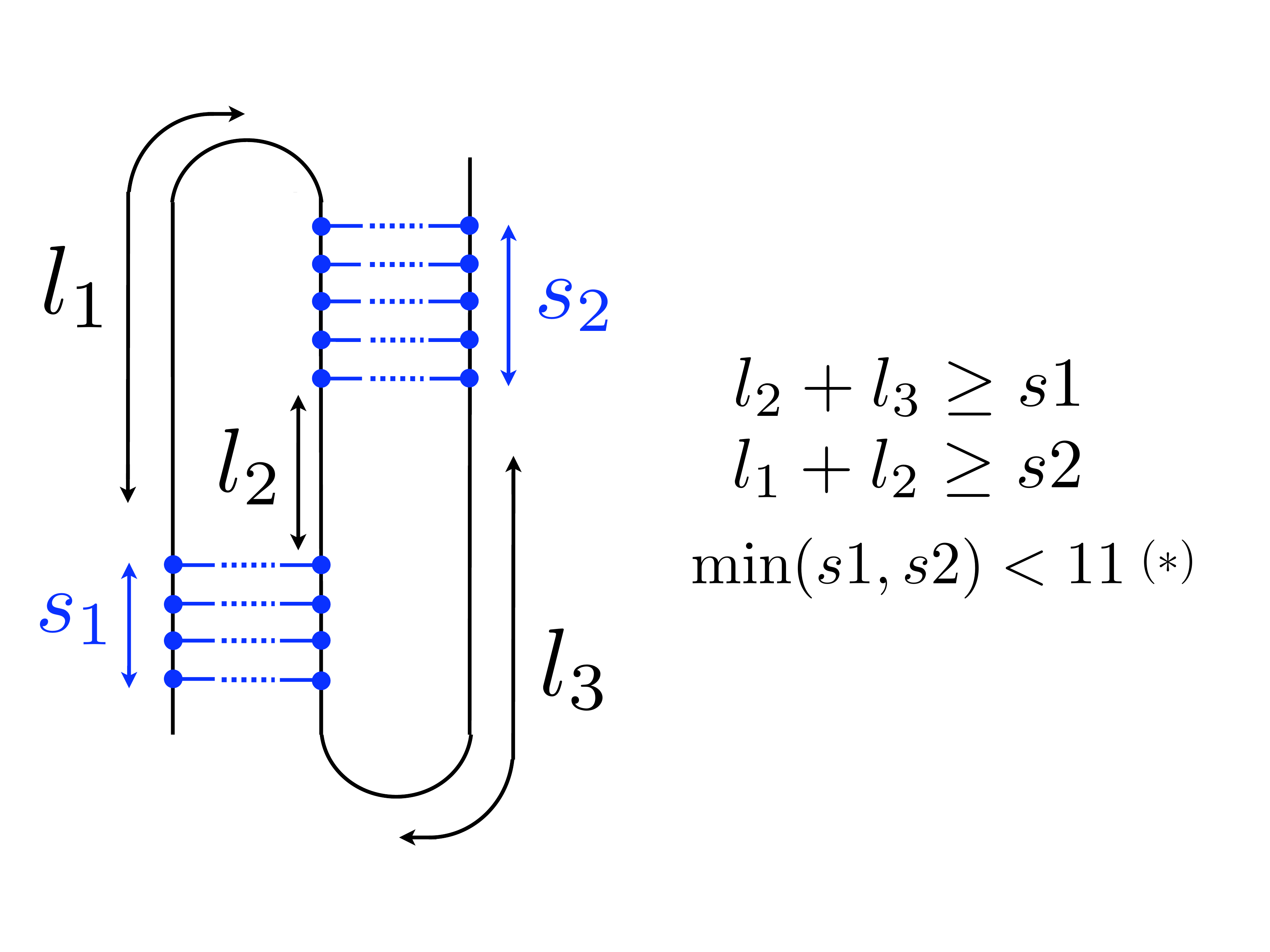}
\end{center}
\caption{\emph{Naive stericity tests used in this work for H pseudoknots. The constraint (*) is used to prevent the formation of real knots.}}
\label{PNH}
\end{figure}

The question thus remains : how to deal with steric constraints ? To our knowledge this is an open question. No clear criteria is known to decide whether a proposed pseudoknot is possible or not. For simple H-pseudoknots, where only two helipoints are involved, it is an easy task : during the generation of the initial graph, it is sufficient to declare two helipoints incompatible if they form a sterically impossible pseudoknot. In this version of TT2NE, we have used a simple test depicted in Fig. \ref{PNH}. However,  this test is not foolproof as TT2NE still wrongly predicts the Wobble pair in the case discussed above.

In this work, despite the lack of an adequate treatment of steric
constraints, for every studied sequence, we have kept the full
initial set of stable helipoints to check how it impacts on the
complexity of the free energy minimization. We also reckoned that
TT2NE cannot be used for sequences larger than 250 bases on a
single standard processor unit, because the large number of
helipoints makes the calculations too long. TT2NE must thus be
seen as a tool for pseudoknot prediction, which somehow proves
that penalizing pseudoknots according to their genus is a relevant
and useful concept. As TT2NE builds RNA folds gradually by adding
helipoints, as soon as a steric constraint verification algorithm
will be available, it will be possible to have an ongoing
procedure that will detect sterically impossible structures and
will stop that branch of the search tree. This procedure will
allow to greatly improve the output of TT2NE, as well as an
important  speeding up of the algorithm, since lots of paths will
no longer be explored. We insist again on the need to tackle the
problem of steric constraints as a necessary condition to
substantially improve the field of pseudoknot prediction.

\acknowledgements{
The authors wish to thank A. Capdepon for setting up the TT2NE server at http://ipht.cea.fr/rna/tt2ne.php.
}

\bibliographystyle{unsrt}
\bibliography{TT2NE_2}

\begin{thebibliography}{10}

\bibitem{book}
D.~Elliot and M.~Ladomery.
\newblock {\em Molecular Biology of RNA}.
\newblock Oxford University Press, 2011.

\bibitem{Tinocco1991}
I.~Tinocco~Jr. and C.~Bustamante.
\newblock {\em Journal of Molecular Biology}, 293, 1991.

\bibitem{nussinov}
R.~Nussinov, G.~Pieczenik, J.R. Griggs, and D.J. Kleitman.
\newblock {Algorithms for loop matchings}.
\newblock {\em SIAM Journal on Applied Mathematics}, 35(1):68--82, 1978.

\bibitem{zuker0}
M.~Zuker and P.~Stiegler.
\newblock {Optimal computer folding of large RNA sequences using thermodynamics
  and auxiliary information}.
\newblock {\em Nucleic Acids Research}, 9(1):133--148, 1981.

\bibitem{mccaskill}
J.S. McCaskill.
\newblock {The equilibrium partition function and base pair binding
  probabilities for RNA secondary structure}.
\newblock {\em Biopolymers}, 29:1105--1119, 1990.

\bibitem{lyngso2}
R.B. Lyngso and C.N.S. Pedersen.
\newblock {RNA pseudoknot prediction in energy-based models}.
\newblock {\em Journal of Computational Biology}, 7(3-4):409--427, 2000.

\bibitem{kuo}
M.Y. Kuo, L.~Sharmeen, G.~Dinter-Gottlieb, and J.~Taylor.
\newblock {Characterization of self-cleaving RNA sequences on the genome and
  antigenome of human hepatitis delta virus.}
\newblock {\em Journal of Virology}, 62(12):4439--4444, 1988.

\bibitem{pleijframe}
E.B. Dam, C.W.A. Pleij, and L.~Bosch.
\newblock {RNA pseudoknots : Translational frameshifting and readthrough on
  viral RNAs}.
\newblock {\em Virus genes}, 4(2):121--136, 1990.

\bibitem{orland}
H.~Orland and A.~Zee.
\newblock {RNA folding and large N matrix theory}.
\newblock {\em Nuclear Physics B}, 620(3):456--476, 2002.

\bibitem{bon}
M.~Bon, G.~Vernizzi, H.~Orland, and A.~Zee.
\newblock {Topological classification of RNA structures}.
\newblock {\em Journal of Molecular Biology}, 379(4):900--911, 2008.

\bibitem{t}
G.~t'Hooft.
\newblock {A planar diagram theory for strong interactions}.
\newblock {\em Nuclear Physics B}, 72(3):461--473, 1974.

\bibitem{mathews1999}
D.H. Mathews, J.~Sabina, M.~Zuker, and D.H. Turner.
\newblock {Expanded sequence dependence of thermodynamic parameters improves
  prediction of RNA secondary structure}.
\newblock {\em Journal of Molecular Biology}, 288(5):911--940, 1999.

\bibitem{Metzler}
D.~Metzler and M.E. Nebel.
\newblock {Predicting RNA secondary structures with pseudoknots by MCMC
  sampling}.
\newblock {\em Journal of Mathematical Biology}, 56(1):161--181, 2008.

\bibitem{ren}
J.~Ren, B.~Rastegari, A.~Condon, and H.H. Hoos.
\newblock {HotKnots: heuristic prediction of RNA secondary structures including
  pseudoknots}.
\newblock {\em RNA}, 11(10):1494--1504, 2005.

\bibitem{mfold}
M.~Zuker.
\newblock {Mfold web server for nucleic acid folding and hybridization
  prediction}.
\newblock {\em Nucleic Acids Research}, 31(13):3406, 2003.

\bibitem{eddy}
E.~Rivas and S.R. Eddy.
\newblock A dynamic programming algorithm for rna structure prediction
  including pseudoknots.
\newblock {\em Journal of Molecular Biology}, 285:2053--2068, 1999.

\bibitem{pseudobase}
F.H.D. Van~Batenburg, A.P. Gultyaev, C.W.A. Pleij, J.~Ng, and J.~Oliehoek.
\newblock {PseudoBase: a database with RNA pseudoknots}.
\newblock {\em Nucleic Acids Research}, 28(1):201, 2000.

\bibitem{mfold_evaluation}
K.J. Doshi, J.J. Cannone, C.W. Cobaugh, and R.R. Gutell.
\newblock {Evaluation of the suitability of free-energy minimization using
  nearest-neighbor energy parameters for RNA secondary structure prediction}.
\newblock {\em BMC bioinformatics}, 5(1):105, 2004.

\bibitem{1rnk}
L.X. Shen, I.~Tinoco, et~al.
\newblock {The structure of an RNA pseudoknot that causes efficient
  frameshifting in mouse mammary tumor virus}.
\newblock {\em Journal of Molecular Biology}, 247(5):963--978, 1995.

\end{thebibliography}

\end{document}